\documentclass{article}
\usepackage{LaThuileFPSpro}
\usepackage[dvips]{graphicx}
\usepackage{float}

\begin{document}
%
\title{
 Neutrino detectors in ice: results and perspectives
  }
\author{
  A. Bouchta       \\
  For the IceCube collaboration$^{*}$\\
  {\em Uppsala University, box 535, S-75121 Uppsala, Sweden}
  }
\maketitle

\baselineskip=11.6pt

\begin{abstract}
 The AMANDA neutrino detector has been in operation at the South Pole for several
 years. A number of searches for extraterrestrial sources of high energy neutrinos have
 been performed. A selection of results is presented in this paper. The much larger
 IceCube detector will extend the instrumented ice volume to a cubic kilometer and 9
 out of 80 planned IceCube strings have been deployed to date. We present the  status for both detectors.
\end{abstract}
\newpage

\section{Introduction}
\subsection{Motivation}

The detection of cosmic neutrinos poses great challenges, but has attractive features
as well.
As opposed to other particles and to electromagnetic radiation,  neutrinos can travel
through huge amounts of matter without being absorbed. They can thus lead to the
discovery of distant production sites hidden from observation with conventional
means. Furthermore, they do not get
deflected by magnetic fields during their journey and so their path points straight
back to their point of origin.
Some known extra-terrestrial sources that  are believed to accelerate particles to
very high energies are prime candidates for neutrino production. In that scenario,
neutrinos are produced as a result of the interaction of accelerated charged particles
with surrounding matter or photons, producing pions, which themselves decay into muons
and neutrinos. The actual observation of neutrinos would thus shed light on the
hadronic nature of the accelerating process.
The energy feeding that process is of gravitational origin and possible candidates are
Active Galactic Nuclei (AGNs),SuperNova Remnants (SNRs) or Gamma Ray Bursters
(GRBs). Models predict a hard energy spectrum $ E^{-2}$, which at high enough
energies has a flux detectable above the background of the much steeper atmospheric
neutrino flux.
 Another possible scheme for neutrino production is one where neutrinos come from the decay of massive relic particles \cite{learned-mannheim}.

\subsection{Detection principle}

High energy neutrinos interact via charged and neutral currents with ordinary matter and produce
charged particles in the process. These particles in turn send out Cherenkov light when
they travel through a medium with a refractive index greater than one.
In order to detect the weakly interacting neutrino, large volumes of matter must  be
instrumented with light sensors. In the cases of AMANDA and IceCube, the ice sheet at the South Pole serves
as the Cherenkov medium. Notice, that the clarity of this ice allows for a sparse instrumentation,
reducing the cost of the detector for a given geometry.
Muon neutrinos produce long range secondary muons (1 km at 200 GeV) sending out light
along their path through the detector.
The direction of the parent neutrino can be directly inferred from that of the muon
since the mean angle between the two particles is proportional to $E^{-0.5}$, with a
value of of $\approx 0.7^\circ$ at 1 TeV.
Electron and tau neutrinos will produce cascades, reconstructed
as point sources using the Cherenkov light, but from which some directional information can be extracted as well.
The main background to  these events comes from atmospheric muons produced in the
atmosphere above the detector, so only events reconstructed as up-going are considered
as signal, since atmospheric muons from this direction get absorbed by the Earth. At PeV energies, the
Earth becomes opaque to neutrinos, including the background made by those produced
in the atmosphere and one has to look for events coming from close to the horizon and above. At
those energies, the $\nu_\tau$ has a characteristic ``Double bang'' signature: after
the first cascade, produced when the $\nu_\tau$ interacts with the ice, the resulting
tau particle decays after hundreds of meters, producing a second cascade. Thus the topologies of the three neutrino flavors become quite distinguishable \cite{topologies}.

\begin{figure}[htb]
\centering\includegraphics[height=10cm]{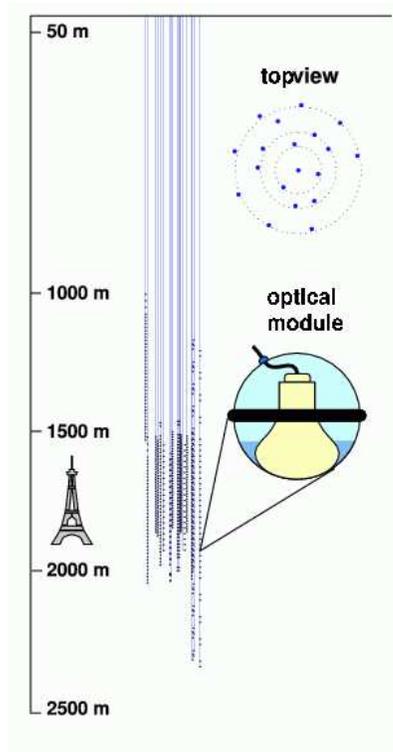}
  \caption{\it
Schematic of the AMANDA detector, with a zoomed in view of one optical module.
    \label{amandadetector} }
\end{figure}
\section{The AMANDA detector}

AMANDA is an array of light sensors installed deep in the 2800 meter thick South Pole ice sheet. Over the years, 19 holes have been melted with a hot-water drill and Optical Modules (OMs) were frozen-in at depths between 1500 and 2000 m below the surface. The OMs are arranged string-wise, with a distance between them of 10 to 20 meters, depending on their location. Each module consists of an 8-inch Hamamatsu photomultiplier, enclosed in a thick glass sphere designed to stand the high pressure of the ice. There are 677 OMs arranged on 19 strings in total, deployed between 1996 and 2000. The signals produced by the photomultipliers travel along cables, connected to a counting house at the surface. After triggering, time- and charge-information is extracted from these pulses and stored for further analysis.
SPASE-II is an air-shower array located at the surface of the ice and operated in coincidence with AMANDA. Muons passing through both detectors have been used to verify the performance of the detector and for position calibration of its OMs.
\begin{figure}[htb]
\centering\includegraphics[width=8cm]{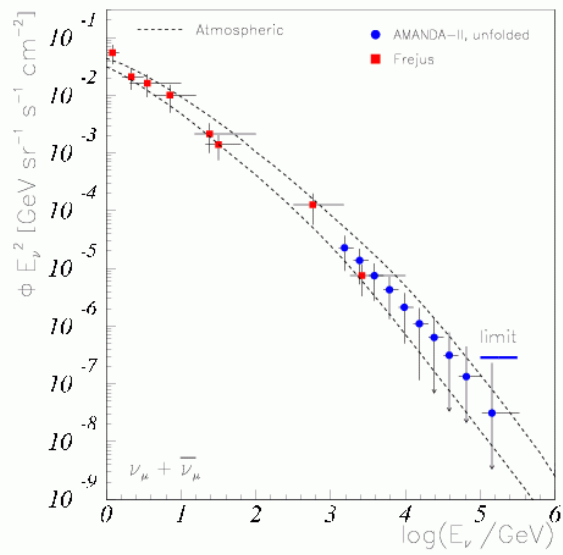}
 \caption{\it The AMANDA and the Frejus measured atmospheric neutrino spectra. The dotted lower line is a parametrization of the horizontal flux and the upper is for a vertical flux\cite{volkova-honda}.\label{unfolded} }
\end{figure}

\section{AMANDA results}
We present here only a selection of results. Several topics are not included here, such as: GRB searches, detection and search for gravitational collapse supernovae, search for exotic particle (Q-balls, nuclearites, monopoles etc.)
\subsection{Atmospheric neutrinos}
Cosmic rays impinging on the atmosphere produce up-going neutrinos which are indistinguishable from extra-terrestrial neutrinos. However, the energy dependence of the flux of the first goes as $E^{-3.7}$ above 1 TeV, in contrast to the flux of the latter which is expected to vary as  $E^{-2}$ and thus decrease much faster with energy.
The energy spectrum of the year 2000 data with 570 neutrino events was extracted using neural net techniques and then unfolded  \cite{neural}. The (preliminary) result is shown in figure \ref{unfolded}, together with the Frejus measurements \cite{daum} at lower energies. AMANDA has performed the first measurements going up to 300 TeV. Both the Frejus and the AMANDA results lie within the horizontal and vertical fluxes parametrized by Volkova ($<100$ GeV) and Honda ($>100$ GeV) \cite{volkova-honda}. The maximum additional flux with an $E^{-2}$ spectrum (at a $90\%$ confidence level) in the range 100-300 TeV is also marked ("limit") on the figure.

\subsection{Point source searches}
\begin{figure}[htb]
\centering\includegraphics[width=10cm]{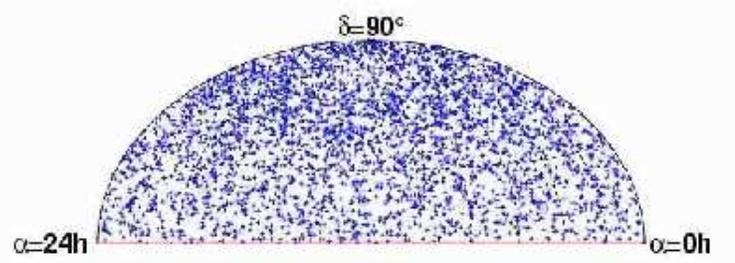}
  \caption{\it Sky-map in celestial coordinates of the northern hemisphere, made using the 3329 neutrino events collected in 2000-2003 by the AMANDA detector. \label{skymap} }
\end{figure}

\begin{figure}[htb]
\centering\includegraphics[width=10cm]{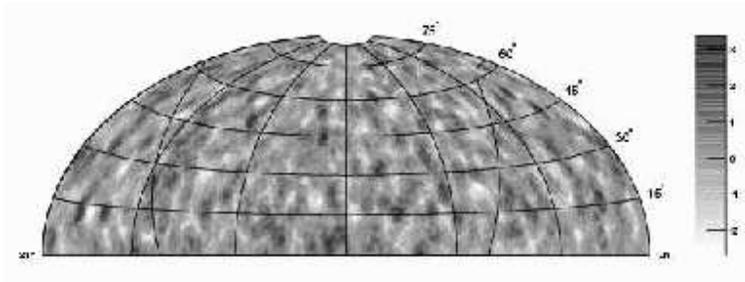}
  \caption{\it Significance map for the excess of events in the North sky. \label{excessmap} }
\end{figure}

3329 $\nu_\mu$ events, collected during 2000-2003, were used in a search for point sources \cite{ackermann}. The sky map of these events is shown in equatorial coordinates in figure \ref{skymap}. A search for a localized source in that map gives Figure \ref{excessmap}, representing the significance of the excess in each direction. The highest value found  in any directions in the whole map corresponds to 3.4$\sigma$. Background simulations tell us that the probability of observing an excess $\ge 3.4\sigma$, taking into account trial factors, is $92\%$.
A selective search for an excess from a list of 33 pre-selected sources does not lead to the discovery of any significant excess either.
\subsection{Diffuse flux search}
\begin{figure}[H]
\centering\includegraphics[width=10cm]{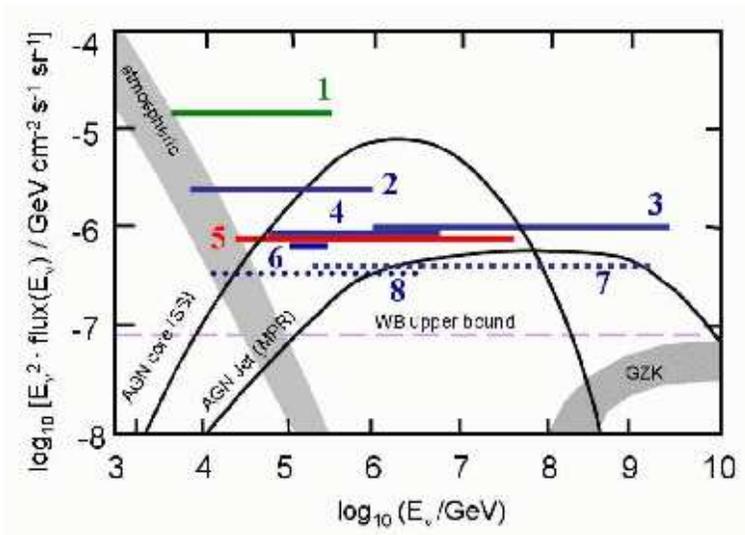}
  \caption{\it Neutrino flux limits from various analyses. All assume neutrino oscillations and full mixing.\label{diffuse} }
\end{figure}
Even if the fluxes from individual objects are too low to be detected separately from each other, the combined flux of many such sources may still be used to offer evidence of the production of extra-terrestrial neutrinos. If that diffuse flux exists, it is expected to have a harder spectrum than that of atmospheric neutrinos and will be exceeding it above a high enough energy.
Figure \ref{diffuse} shows the limits set by AMANDA and other detectors on diffuse neutrinos: the limit labeled (1) is the one set by MACRO \cite{macro}, (2) is from the 10-string stage of AMANDA in 1997 using $\nu_\mu$ \cite{ahrens-b10}, (3) comes from the same data as (2) but the analysis combines all flavors of UHE $\nu$s\cite{uhe}. (4) is the AMANDA 2000 all-flavor cascade limit\cite{cascade-limit}, (5) is the Baikal limit from their 1998-2000 cascade analysis\cite{baikal}. Lines (6),(7) and (8)  are all preliminary results from 2000 data. The line for the limit (6) comes the $\nu_\mu$ analysis\cite{munich}, (7) the sensitivity for all the flavors of UHE muons\cite{lisa}
 and (8) the 2000-2003 $\nu_\mu$ sensitivity\cite{hodges}. We assume full mixing, so that in analysis (1), (2), (6) and (8), the limits for $\nu_\mu$ are shown multiplied by 3. The line marked "AGN core (SS)" is the expected flux of diffuse neutrinos according to the model described in \cite{SS} and the one marked "AGN jet (MPR)" refers to the model in \cite{MPR}. Both models have been scaled by 1.5 to account for oscillations. The Waxman-Bahcall upper limit ("WB")\cite{WB} is also indicated on the figure.

\subsection{Dark matter}
\begin{figure}[htb]
\centering\includegraphics[width=6.3cm]{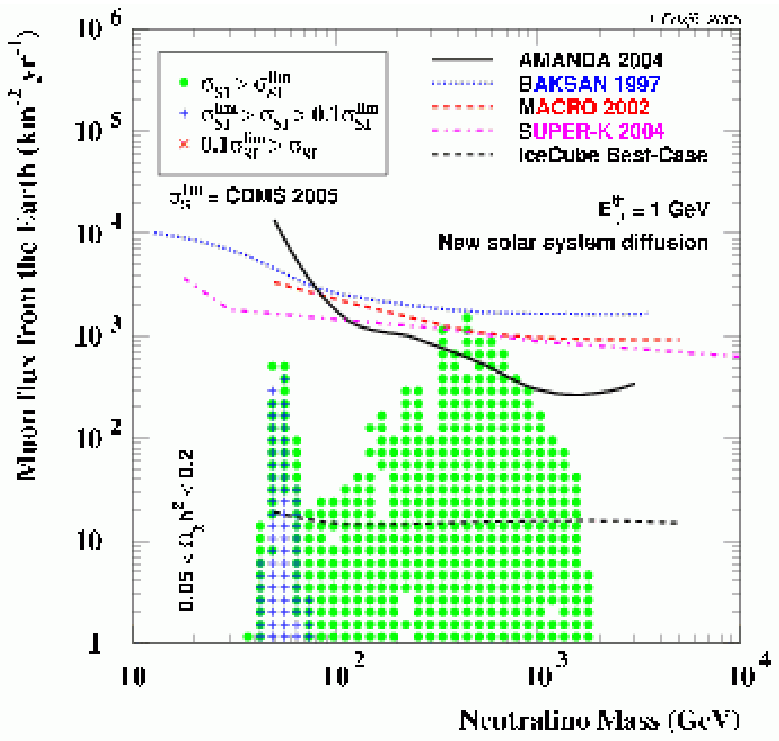}\includegraphics[width=6cm]{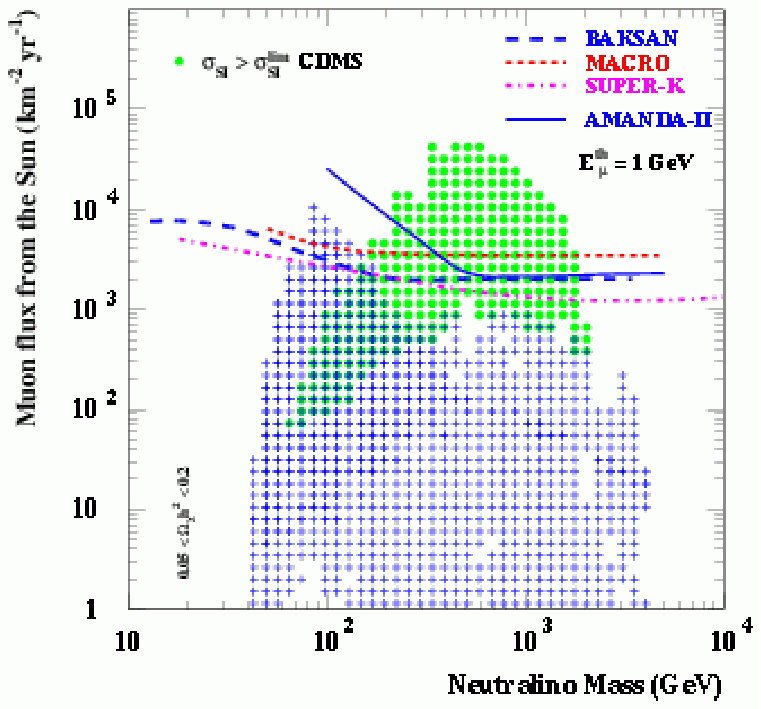}
  \caption{\it Limits at $90\%$ C.L. on the muon flux from neutrinos produced in neutralino annihilations in the center of the Earth (left) and the center of the Sun (right). The dots show models disfavored by direct search results from CDMS and the crosses correspond to predictions from different MSSM models. \label{wimp} }
\end{figure}
A prime candidate WIMP particle to make up for the cold dark matter of the Universe is the lightest supersymmetric particle, the neutralino. An indirect observation of these particles can be made using AMANDA in the following scenario. Neutralinos lose their energy in collisions with matter and get gravitationally trapped inside massive objects, where they annihilate with each other, emitting neutrinos in that process.  The closest and in that sense most interesting sources of accumulation are the center of the Earth and the Sun. The limits at $90\%$ C.L. using 1997-1999 data for the Earth \cite{earth-wimps} and using 2000 data for the Sun\cite{sun-wimps}, are shown on Figure \ref{wimp}, together with the limits for other experiments and projected results for IceCube.

\section{The IceCube detector}
\begin{figure}[htb]
\centering
\includegraphics[width=7cm]{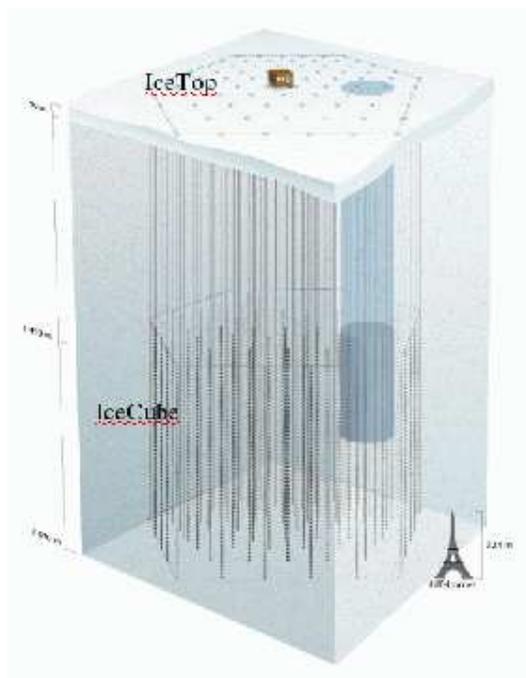}
   \caption{\it A schematic view of the IceCube detector. AMANDA can be seen enclosed inside the IceCube volume and the IceTop array at the surface of the ice. \label{icecube-detector} }
\end{figure}
\begin{figure}[htb]
\centering
\includegraphics[width=10cm]{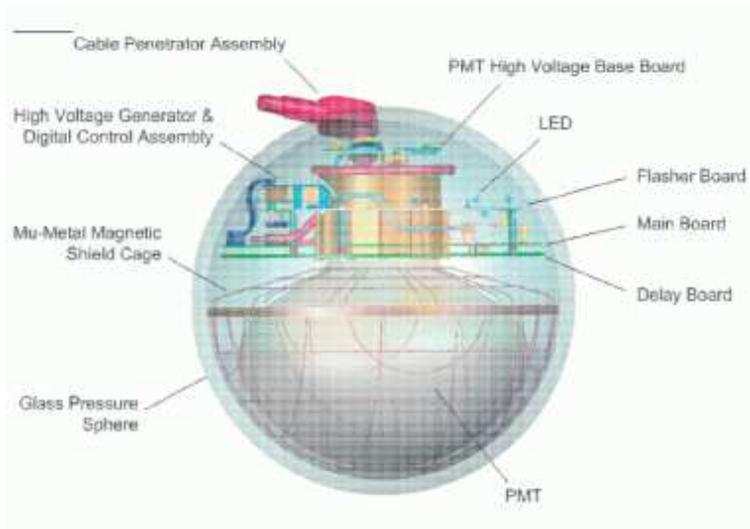}
  \caption{\it A Digital Optical Module (DOM). \label{dom}}
\end{figure}
IceCube follows the successful concept demonstrated by AMANDA, but on a larger scale and with more advanced hardware.
With IceCube, we enter the era of cubic kilometer sized neutrino telescopes. At completion, it will consist of 80 strings arranged on a regular hexagon and with 60 Digital Optical Modules (DOMs) each (see Figure \ref{icecube-detector}). The instrumented region will  be between 1450 m and 2450 m below the surface and the hexagonal pattern will span over a $\rm km^2$ area. The distance between DOMs within a string is 17 m and the distance between the strings is 125 m, with the existing AMANDA detector embedded inside the IceCube volume. Strings will be added each year until completion but the detector is taking data since the start.
With it's large size, IceCube should observe $\nu$ events with energies up to $10^9 GeV$ and collect many contained events.
An air shower array, IceTop, is deployed at the surface of the ice above IceCube. It consists of 160 Cherenkov tanks located close to the 80 holes. The tanks are grouped in pairs ('stations') near a hole and are filled with water which is then  allowed to freeze around two DOMs inside. IceTop will serve several purposes: cosmic ray studies up to energies up to $10^{17} \rm eV$), search for a possible second "knee" for heavy particles like iron, calibration of the IceCube DOMs using coincident atmospheric muons, vetoing of highly energetic background events originating in the atmosphere.

\subsection{Design and construction}
The DOM (Figure \ref{dom}) is the main component of the detector. It contains a 10-inch Hamamatsu photomultiplier tube (PMT), enclosed in a glass vessel. An integrated circuit (Analog Transient Waveform Digitizer, or ATWD) records the PMT pulse with three different gains (x1/4, x2 and x16). The ATWD records 128 samples with a rate of 3.3 ns/sample. The same pulse is also digitized by a 40MHz FADC. The linear dynamic range is 400 photo-electrons (p.e) in 15 ns and the integrated dynamic range is more than 5000 p.e. in $2\mu \rm s$. Neighboring DOMs are connected to each other and the ATWD is read out after a local triggering condition is met. Each waveform is time-stamped locally. The main board contains an FPGA that can be re-programmed from a computer at the surface.
The DOMs communicate with the counting house with twisted pair copper cables and receive their power through the same cables. Each DOM has its own high-voltage generator. Since the pulses are sent  digitally, the PMT amplification can be kept low at $0.5-1.0 \times 10^7$. This benefits both the dynamic range  and the aging of the PMTs.\\
The DOMs are deployed in holes melted with a hot-water drill. A new 5 MW drill, the Enhanced Hot Water Drill (EHWD) was developed for IceCube and put into operation in January 2005. This new construction is designed to drill 2500 m deep holes with a 60 cm diameter in less than 40 h, much faster than the AMANDA drill.
The actual deployment of the DOMs takes less than 12 hours. With the EHWD, IceCube will be able to deploy 14 strings/year during the November to mid-February period when the South Pole station is open for construction work.
Simulations of the detector have been made \cite{ahrens} to predict its capabilities.
The effective area  is about $0.8 \rm km^{2}$ for 1 TeV muons  The angular resolution is expected to be $0.7^{\circ}$ at $10\rm TeV$.
\subsection{Icecube status}
\begin{figure}[H]
\centering
\includegraphics[width=8cm]{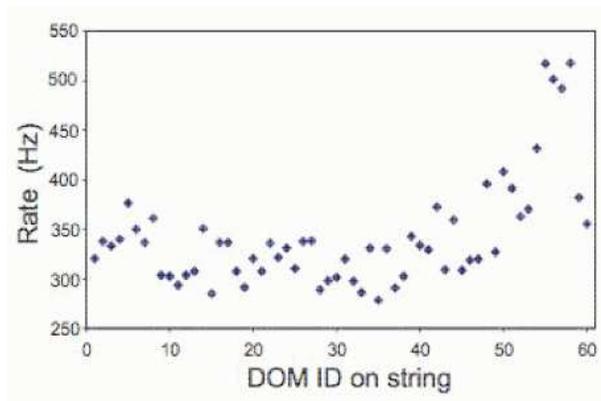}
  \caption{\it The noise rate of individual DOMs on the first string as a function of DOM ID (depth increases with DOM ID). The rates are measure including a dead-time of $51\mu\rm s$. \label{noise}}
\end{figure}
\begin{figure}[htb]
\centering
\includegraphics[width=12cm]{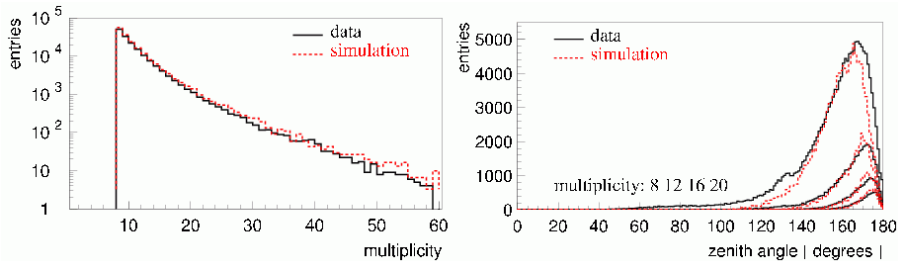}
  \caption{\it Comparison of data taken with string 21, and simulations of atmospheric muons. The left figure shows the number of hit DOMs. The right figure shows the reconstructed zenith angle for different multiplicities. \label{comparison}}
\end{figure}
\begin{figure}[H]
\centering
\includegraphics[width=3cm]{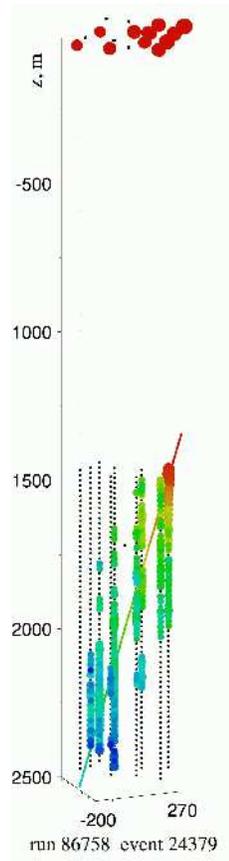}
  \caption{\it An event acquired with the current nine-string IceCube detector. A shower passes through IceTop and the muon component leaves a track in IceCube. \label{event}}
\end{figure}

A first string, labelled '21' was deployed in February 2005, together with four IceTop stations, totalling 76 DOMs together\cite{icrc}.  Since then, all DOMs of string 21 have been taking data and behaving according to expectation. In particular, the time calibration of quartz oscillator in each DOM has been studied. These oscillators are synchronized to a master GPS clock every 3.5 seconds, by sending down pulses from the surface to the DOMs. The calibration procedure yields a time precision of 3 ns.
Another measurement is the dark noise rate of the DOMs, which is around 700 Hz. With after-pulse suppression (implemented with an artificial dead-time of $51 \mu \rm s$), this rate goes down to about 350 Hz, see Figure \ref{noise}. In the same way as AMANDA, IceCube will be able to detect the burst of low energy neutrinos that accompany a supernova explosion by measuring an excess rate above the dark noise in all its sensors. A low dark noise improves the sensitivity for such a search.
String 21 has also been used to reconstruct atmospheric muons. The resulting multiplicities and zenith angles agree with simulations (see Figure \ref{comparison} for the comparison).
In January 2006, the installation of eight more  strings and 12 additional IceTop
stations was completed. An event passing through IceTop and the nine IceCube strings is
shown in Figure \ref{event}.  These new strings total 540 DOMs (and IceTop 64 DOMs)
which cover a larger volume than AMANDA's 677 modules. $99\%$ of the DOMs have survived
the deployment and re-freezing period and IceCube took data for several months at an average rate of 79Hz (with a trigger condition of 8 DOMs hit in a window of $2\mu\rm s$).

\section{Conclusions}
We have presented a selection of the AMANDA results from data taken over the past ten years. The transition from AMANDA to the much larger IceCube detector is under way . The performance of the first IceCube string is as expected. The experience gained in deploying eight additionalstrings with new equipment bodes well for the higher rate of deployment in coming years.
\section{Acknowledgements}

 We acknowledge the support from the following agencies: National Science Foundation-Office of Polar Program, National Science Foundation-Physics Division, University of Wisconsin Alumni Research Foundation, Department of Energy, and National Energy Research Scientific Computing Center (supported by the Office of Energy Research of the Department of Energy), the NSF-supported TeraGrid system at the San Diego Supercomputer Center (SDSC), and the National Center for Supercomputing Applications (NCSA); Swedish Research Council, Swedish Polar Research Secretariat, and Knut and Alice Wallenberg Foundation, Sweden; German Ministry for Education and Research, Deutsche Forschungsgemeinschaft (DFG), Germany; Fund for Scientific Research (FNRS-FWO), Flanders Institute to encourage scientific and technological research in industry (IWT), Belgian Federal Office for Scientific, Technical and Cultural affairs (OSTC); the Netherlands Organisation for Scientific Research (NWO); M. Ribordy acknowledges the support of the SNF (Switzerland); J. D. Zornoza acknowledges the Marie Curie OIF Program (contract 007921).

%
%
$^{*}$
{\scriptsize
\begin{sloppypar}

\noindent
 A.~Achterberg$^{30}$,
 M.~Ackermann$^{32}$,
 J.~Adams$^{10}$,
 J.~Ahrens$^{20}$,
 K.~Andeen$^{19}$,
 D.~W.~Atlee$^{28}$,
 J.~N.~Bahcall$^{24,a}$,
 X.~Bai$^{22}$,
 B.~Baret$^{8}$,
 M.~Bartelt$^{12}$
% ...

\end{sloppypar}

\vspace*{0.5cm}

{\it
\noindent
  (1) Dept. of Physics and Astronomy, University of Alaska Anchorage, 3211 Providence Dr., Anchorage, AK 99508, USA \newline
 (2) CTSPS, Clark-Atlanta University, Atlanta, GA 30314, USA \newline
 (3) Dept. of Physics, Southern University, Baton Rouge, LA 70813, USA \newline
 (4) Dept. of Physics, University of California, Berkeley, CA 94720, USA \newline
 (5) Institut f\"ur Physik, Humboldt Universit\"at zu Berlin, D-12489 Berlin, Germany \newline
 (6) Lawrence Berkeley National Laboratory, Berkeley, CA 94720, USA \newline
 (7)  Universit\'e Libre de Bruxelles, Science Faculty CP230, B-1050 Brussels, Belgium \newline
 (8) Vrije Universiteit Brussel, Dienst ELEM, B-1050 Brussels, Belgium \newline
 (9) Dept. of Physics, Chiba University, Chiba 263-8522 Japan \newline
 (10)  Dept. of Physics and Astronomy, University of Canterbury, Private Bag 4800, Christchurch, New Zealand \newline
(11)Dept. of Physics, University of Maryland, College Park, MD 20742, USA \newline
(12)  Dept. of Physics, Universit\"at Dortmund, D-44221 Dortmund, Germany \newline
(13)   Dept. of Subatomic and Radiation Physics, University of Gent, B-9000 Gent, Belgium \newline
(14)  Max-Planck-Institut f\"ur Kernphysik, D-69177 Heidelberg, Germany \newline
(15)   Dept. of Physics and Astronomy, University of California, Irvine, CA 92697, USA \newline
(16)   Dept. of Physics and Astronomy, University of Kansas, Lawrence, KS 66045, USA \newline
(17)  Blackett Laboratory, Imperial College, London SW7 2BW, UK \newline
(18)  Dept. of Astronomy, University of Wisconsin, Madison, WI 53706, USA \newline
(19)  Dept. of Physics, University of Wisconsin, Madison, WI 53706, USA \newline
(20)   Institute of Physics, University of Mainz, Staudinger Weg 7, D-55099 Mainz, Germany \newline
 (21) University of Mons-Hainaut, 7000 Mons, Belgium \newline
 (22) Bartol Research Institute, University of Delaware, Newark, DE 19716, USA \newline
 (23) Dept. of Physics, University of Oxford, 1 Keble Road, Oxford OX1 3NP, UK \newline
 (24) Institute for Advanced Study, Princeton, NJ 08540, USA \newline
 (25) Dept. of Physics, University of Wisconsin, River Falls, WI 54022, USA \newline
 (26) Dept. of Physics, Stockholm University, SE-10691 Stockholm, Sweden \newline
 (27)  Dept. of Astronomy and Astrophysics, Pennsylvania State University, University Park, PA 16802, USA \newline
 (28)  Dept. of Physics, Pennsylvania State University, University Park, PA 16802, USA \newline
 (29)  Division of High Energy Physics, Uppsala University, S-75121 Uppsala, Sweden \newline
 (30)  Dept. of Physics and Astronomy, Utrecht University/SRON, NL-3584 CC Utrecht, The Netherlands \newline
  (31) Dept. of Physics, University of Wuppertal, D-42119 Wuppertal, Germany \newline
 (32) DESY, D-15735 Zeuthen, Germany \newline
 (a) Deceased \newline
 (b) On leave of absence from Universit\`a di Bari, Dipartimento di Fisica, I-70126, Bari, Italy \newline
 (c) Affiliated with Dept. of Chemistry and Biomedical Sciences, Kalmar University, S-39182 Kalmar, Sweden \newline
(d) Now at RWTH Aachen University, D-52056 Aachen, Germany \newline
 (e) Affiliated with IFIC (CSIC-Universitat de Val\`encia), A. C. 22085, 46071 Valencia, Spain

}}


\begin{thebibliography}{99}
\bibitem{learned-mannheim} J.G. Learned and K. Mannheim, Ann. Rev. Nucl. Part. Sci. 2000.50:679
\bibitem{SS} F.W.~Stecker and M.H.~Salamon, Space. Sci. Rev.{\bf 75} (1996) 341
\bibitem{MPR} K.~Mannheim, R.~Protheroe and J.~Rachen, Phys. Rev.{\bf D63} (2001) 023003
\bibitem{WB} E.~Waxman and J.~Bahcall, Phys. Rev.{\bf D59} (1999) 023002

\bibitem{topologies} J.G.~Learned and S.~Pakvasa, Astropart. Phys. {\bf3} (1995) 267
\bibitem{neural} K. Woschnagg {\it et al.} Nucl. Phys. B. (Proc. Suppl.) {\bf 142} (2005) 343-350

\bibitem{daum} K. Daum {\it et al.} Zeitschrift f\"ur Physik {\bf C 266} (1995) 417

\bibitem{volkova-honda} L.V. Volkova, Sov. J. Nucl. Phys. {\bf 31} (1980); M. Honda {\it et al.}, Phys. Rev. D {\bf 52}  (1995) 4985

\bibitem{ackermann} M. Ackermann for the IceCube Collaboration, ICRC Pune 2005; astro-ph/0509330

\bibitem{macro} M. Ambrosio {\it et al.}, Astropart. Phys. {\bf 19} (2003) 1


\bibitem{ahrens-b10} J. Ahrens {\it et al.}, Phys. Rev. Let. {\bf 90} (2003) 251101


\bibitem{uhe} M. Ackermann {\it et al.}, Astropart. Phys. {\bf 22} (2005) 339

\bibitem{cascade-limit} M. Ackermann {\it et al.}, Astropart. Phys. {\bf 22} (2004) 127

\bibitem{baikal} V. Aynutdinov  {\it et al.}, Astropart. Phys. {\bf 25} (2006) 140

\bibitem{munich} K. M\"unich for the IceCube callaboration, ICRC Pune 2005; astro-ph/0509330

\bibitem{lisa} L. Gerhardt for the IceCube collaboration, ICRC Pune 2005; astro-ph/0509330

\bibitem{hodges} J. Hodges  for the IceCube callaboration, ICRC Pune 2005; astro-ph/0509330

\bibitem{earth-wimps} A. Achterberg  {\it et al.}, accepted for publication in  Astropart. Phys.
\bibitem{sun-wimps} M. Ackermann {\it et al.},  Astropart. Phys. {\bf 24} (2006) 459

\bibitem{ahrens} J. Ahrens {\it et al.},  Astropart. Phys. {\bf 20} (2004) 507
\bibitem{icrc} A.~Achtenberg et al., astro-ph/0604450

\end{thebibliography}
\end{document}